\documentclass{elsart}

\usepackage{natbib}
\usepackage{graphicx}

\begin{document}

\runauthor{Felipe Valencia, Aldo H. Romero, Miguel Kiwi, Ricardo
 Ram\'{\i}rez and Alejandro Toro-Labb\'e}

\begin{frontmatter}
  
  \title{Internal Rotation of Disilane and Related Molecules: a
    Density Functional Study\thanksref{Fondecyt}}

\author[PUCFF]{Felipe Valencia}\ead{fvalencia@bethe.fis.puc.cl}\ 
\author[PUCFF]{Aldo H. Romero}\ead{ahromero@puc.cl}\
\author[PUCFF]{Miguel Kiwi}\ead{mkiwi@puc.cl}
\centering{
\author[PUCFF]{Ricardo Ram\'{\i}rez}\ead{rramirez@puc.cl}\
\author[PUCFQ]{Alejandro Toro-Labb\'e}\ead{atola@puc.cl}
}

\address[PUCFF]{Facultad de F{\'\i}sica, Universidad Cat{\'o}lica de Chile\\
  Casilla 306, Santiago, CHILE 6904411} 

\address[PUCFQ]{Facultad de   Qu\'{\i}mica, Universidad Cat{\'o}lica de Chile\\
  Casilla 306, Santiago, CHILE 6904411}

\thanks[Fondecyt]{Supported by the \textit{Fondo Nacional de Desarrollo Cient{\'\i}fico y
Tecnol{\'o}gico} (FONDECYT, Chile) under grants \#8990005, 1020534 and 1010988.}

\begin{abstract}
  
  DFT calculations performed on Si$_2$H$_6$, Si$_2$F$_6$,
  Si$_2$Cl$_6$, and Si$_2$Br$_6$ are reported. The evolution of the
  energy, the chemical potential and the molecular hardness, as a
  function of torsion angle, is studied. Results at the
  DFT-B3LYP/6-311++G** level show that the molecules always favor the
  stable staggered conformations, with low but significant energy
  barriers that hinder internal rotation.  The chemical potential and
  hardness of Si$_2$H$_6$ remains quite constant as the sylil groups
  rotate around the Si-Si axis, whereas the other systems exhibit
  different degrees of rearrangement of the electronic density as a
  function of the torsion angle. A qualitative analysis of the
  frontier orbitals shows that the effect of torsional motion on
  electrophilic attack is negligible, whereas this internal rotation
  may generate different specific mechanisms for nucleophilic attack.

\end{abstract}

\end{frontmatter}

\section{Introduction}
\label{sec:intro}

\begin{figure}[ht]
\centering
\resizebox{\columnwidth}{!}{\includegraphics{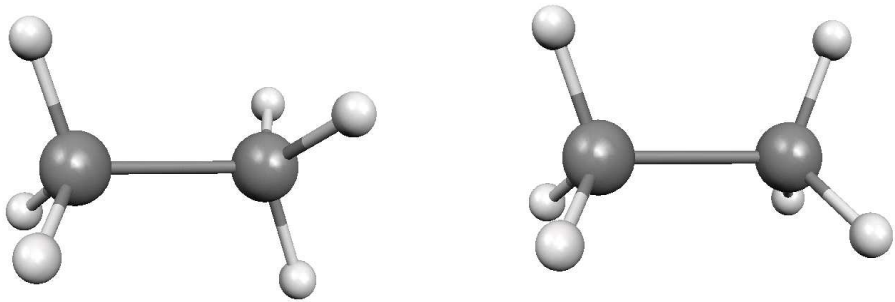}}
\caption{Staggered (left) and eclipsed (right)
  conformations. The rotation angle $\alpha = 0^\circ$ for the
  staggered and $\alpha = 60^\circ$ for the eclipsed conformations,
  respectively.}
\label{fig:molecule}
\end{figure}

Recently disilane has attracted attention~\cite{poph01,cho97} due
to its importance in the production of silicon based semiconductor
devices. Its geometry is quite similar to ethane, which is the
best known and most widely studied
example~\cite{jones,march,loudon,mcmurry} of simple molecules with
properties that markedly depend on the rotation of a group of
atoms around one or more internal bonds, going through stable and
unstable conformations as a full 360$^\circ$ rotation is executed.
In particular, the central C-C bond of ethane is a threefold
symmetry axis. Thus, as one of the two methyl groups rotates
around this axis the molecule goes through (stable) staggered and
(unstable) eclipsed conformations (see Fig.~\ref{fig:molecule}).
The preferred staggered structure is attributed to steric
effects~\cite{jones,march,loudon,mcmurry,pop,badenhoop}, more
precisely to increased repulsion between electrons in bonds that
are drawn closer together~\cite{mcmurry}.

On the other hand, the fundamental processes in the disilane
decomposition on silicon surfaces are relevant to the
understanding and optimization of the growth of epitaxial silicon
on silicon substrates.  The morphological parameters of the
eclipsed and staggered silane conformations were recently
calculated by Pophristic and co-workers~\cite{poph01}. They
concluded that the origin of the eclipsed to staggered relaxation
is related to the preferential hyperconjugative stabilization
(meaning energy stabilization through electron excitation to a
delocalized state). This charge delocalization changes the
electronic properties of the molecule, as a function of the
conformation it adopts.

When a reaction moves forward along the reaction coordinate, a
redistribution of the ground--state electron density takes place, and
the resulting energy change can be understood in terms of the response
of the system to variations of the total number of electrons $N$, and
of the external $v(\vec{r})$ potential~\cite{parr88}. Density
functional theory (DFT)~\cite{parr88,dreizler} has been quite
successful in providing a theoretical basis for qualitative chemical
concepts like chemical potential ($\mu$) and hardness ($\eta$), which
describe the response of the system when $N$ is varied for a fixed
$v(\vec{r})$~\cite{parr88}. In DFT $\mu$ is the Lagrange multiplier
associated with the normalization constraint that requires
conservation of the number of electrons $N$. Classical structural
chemistry is recovered with the identification of $\mu$ as minus the
electronegativity ($\mu = -\chi$), a well known and well established
quantity. Definitions of $\mu$ and $\eta$, two global electronic
properties that are implied in the reactivity of molecular systems,
were given by Parr {\it et al.}~\cite{parrdonn} and Parr and
Pearson~\cite{ppierson,pierson97}, respectively. The application of
DFT concepts to the analysis of chemical reactions is better
appreciated with the help of the principle of maximum hardness (PMH),
that asserts that molecular systems reach equilibrium tending towards
states with the highest hardness~\cite{pmh1,pmh2}. In this
context the PMH can also be helpful in identifying transition
states where minimum values of $\eta$ are expected~\cite{atl01}.

The main purpose of this paper is to provide a detailed report on
the geometric and electronic structure of disilane (Si$_2$H$_6$)
and the family of closely related molecules Si$_2$X$_6$, where X =
F, Cl and Br, as well as the implications this structure has on
the molecular properties.  We focus our attention on the changes
that are induced on the energy and molecular properties as the
molecular conformation periodically changes from staggered to
eclipsed and back to staggered through rotation with respect to
the Si-Si bond of the SiX$_3$ group of Si$_2$X$_6$ (X=H,F,Cl,Br).

This paper is organized as follows: after this introduction we discuss
technical aspects of our calculation in Sec.~\ref{sec:comput}, discuss
the molecular geometry in Sec.~\ref{sec:geom}, the electronic energy
profiles and rotational barriers in Sec.~\ref{sec:energy} and the
chemical potential and the hardness in Sec.~\ref{sec:chem_pot}. In
Sec.~\ref{sec:reactivity} we present a qualitative analysis of the
chemical reactivity of silanes and finally, we close the paper in
Sec.~\ref{sec:concl} drawing conclusions.

\section {Computational Details}
\label{sec:comput}

The calculations reported in this paper were performed using the
Gaussian 98~\cite{gaus98} package. The results we report were
obtained implementing the B3LYP--DFT method and corresponds to the
B3LYP/6-311++G** level. A lower level B3LYP/6-31G{*} calculation
was tested for comparison purposes.  For the staggered and
eclipsed conformations, and in order to test the DFT results,
MP2/6-311++G** calculations were also carried out to check the
quality of the B3LYP results.

\section{Geometry}
\label{sec:geom}

Constrained geometrical optimization was performed varying the
dihedral angle $\alpha$, defined as the rotation angle of the
silyl group of Si$_2$H$_6$ (or the three X atoms in Si$_2$X$_6$),
located at one end of the Si-Si bond, relative to the same three
atoms at the other end of this bond (see the illustration in
Fig.~\ref{fig:molecule}). The angle $0 \le \alpha < 60^{\circ}$,
with $\alpha = 0$ defined as the staggered conformation and
$\alpha = 60^{\circ}$ as the eclipsed one, is varied in steps of
10$^{\circ}$.

\begin{table}
\small {\centering \begin{tabular}{|c|c|c|c|} \hline Molecule&
 B3LYP/6-31G{*}&
 B3LYP/6-311++G{*}{*}&
 Experiment\\
\hline Si$_2$H$_6$ (staggered) & & & \\
 d(Si-Si)&
 2.350&
 2.354&
 2.331~\cite{cho97}\\
 d(Si-H)&
 1.489&
 1.487&
 1.492~\cite{cho97}\\
 $\angle$ (SiSiH) &
 110.4&
 110.6&
 110.3~\cite{cho97}\\
 Si$_2$H$_6$ (eclipsed)& & & \\
 d(Si-Si)&
 2.360&
 2.366&
\\
 d(Si-H)&
 1.489&
 1.487&
\\
 $\angle$ (SiSiH)&
 110.8&
 110.6&
\\
\hline Si$_2$F$_6$ (staggered) & & & \\
 d(Si-Si)&
 2.319&
 2.336&
 2.317~\cite{cho97}\\
 d(Si-F)&
 1.593&
 1.598&
 1.564~\cite{cho97}\\
 $\angle$ (SiSiF) &
 110.5&
 110.7&
 110.3~\cite{cho97}\\
 Si$_2$F$_6$ (eclipsed) & & & \\
 d(Si-Si)&
 2.326&
 2.341&
\\
 d(Si-F)&
 1.592&
 1.598&
\\
 $\angle$ (SiSiF) &
 110.7&
 110.7&
\\
\hline Si$_2$Cl$_6$ (staggered) & & & \\
 d(Si-Si)&
 2.355&
 2.354&
 2.320\\
 d(Si-Cl)&
 2.060&
 2.056&
 2.002\\
 $\angle$ (SiSiCl) &
 109.7&
 109.6&
\\
 Si$_2$Cl$_6$ (eclipsed) & & & \\
 d(Si-Si)&
 2.377&
 2.378&
\\
 d(Si-Cl)&
 2.059&
 2.056&
\\
 $\angle$ (SiSiCl) &
 110.0&
 109.9&
\\
\hline Si$_2$Br$_6$ (staggered)& & & \\
 d(Si-Si)&
 2.335&
 2.368&
\\
 d(Si-Br)&
 2.211&
 2.232&
\\
 $\angle$ (SiSiBr) &
 108.7&
 109.2&
\\
 Si$_2$Br$_6$ (eclipsed)& & & \\
 d(Si-Si)&
 2.356&
 2.405&
\\
 d(Si-Br)&
 2.209&
 2.232&
\\
 $\angle$ (SiSiBr) &
 108.7&
 110.0&
 \\
\hline
\end{tabular}\par}

\caption{Calculated geometries of \protect\protect  Si$_2$H$_6$,
  \protect \protect   \protect\protect Si$_2$F$_6$ \protect
  \protect  \protect\protect Si$_2$Cl$_6$ \protect \protect
and \protect\protect Si$_2$Br$_6$\protect \protect}

\label{table:parameters}
\end{table}

\normalsize The structural parameters obtained for the staggered
conformation are listed and compared to experimental data
-whenever the latter is available- in
Table~\ref{table:parameters}. It is apparent that for the lighter
molecules ({\it i.e.} Si$_2$H$_6$ and Si$_2$F$_6$) B3LYP/6-31G{*}
yields better agreement with experimental values than
B3LYP /6-311++G**, while the larger basis set 6-311++G{*}{*} fares
better for Si$_2$Cl$_6$. We expect the same to hold for
Si$_2$Br$_6$ (a molecule that has not yet been synthesized) since
heavier atoms require larger basis sets for a proper description.
Thus, in what follows below, our comparisons with experiment are
based upon the results of B3LYP/6-311++G{*}{*}.

\begin{figure}
  \centering
  \resizebox*{!}{7cm}{\rotatebox{270}{\includegraphics{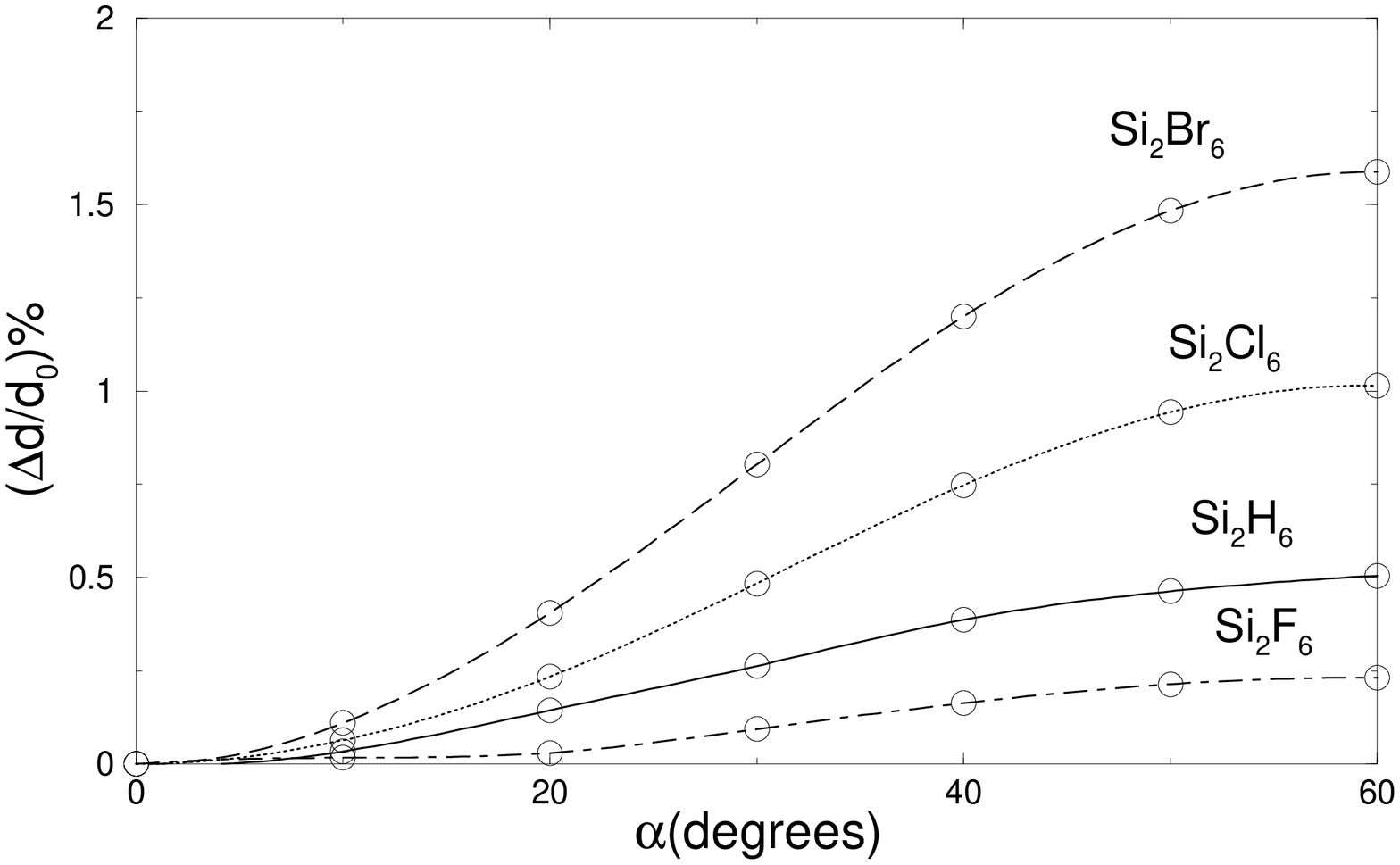}
      }} \par{} \ \caption{Change of the Si-Si distance for the four
    molecules studied, in percentages.  The open circles are the
    calculated points and the lines are guides to the eye. }
\label{fig:dist_vs_angle}
\end{figure}

Rotation of one sylil group with respect to the staggered
conformation is accompanied by a significant change in the Si-Si
distance (see Fig.~\ref{fig:dist_vs_angle}) while the Si-X (X=H,
F, Cl and Br) distance remains almost unchanged (see
Table~\ref{table:parameters}). In fact, Si$_2$Br$_6$ displays the
largest deformation, which amounts to about 1.59\%, while
Si$_2$F$_6$ undergoes a tiny elongation of only 0.23\%~.

The angle $\angle$(SiSiX), between the Si axis and the
X-atoms, exhibits a small but systematic increase as a function of
$\alpha$. Again, this change is largest for Si$_2$Br$_6$
(approximately 0.66\%) and smallest for Si$_2$F$_6$ (approximately
0.04\%).

\section{Energy Profiles and Rotational Barriers}
\label{sec:energy}

\begin{figure}
\centering
\resizebox*{!}{7cm}{\rotatebox{270}{\includegraphics{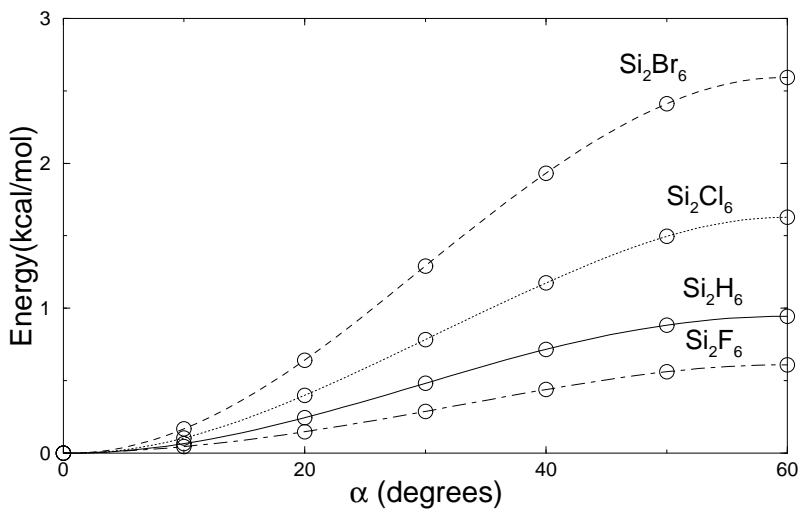}
    }} \par{}
\caption{Electronic energy as function of torsion angle $\alpha$. The
  open circles are the calculated points and the lines are guides to
  the eye.}
\label{fig:energy_vs_angle}
\end{figure}

Fig.~\ref{fig:energy_vs_angle} shows the evolution of the total
energy for each molecule studied, measured with respect to the
total energy in the staggered configuration. In each case,
staggered conformation is of minimum energy and eclipsed
conformation presents maximum energy.

It is also evident from Figs.~\ref{fig:dist_vs_angle} and
\ref{fig:energy_vs_angle} that the energy follows the same trend as
the Si-Si distance along the torsional angle. Clearly, the torsional
potential energy can be understood in terms of the structural changes
of the molecule undergoes as $\alpha$ is varied.  Si$_2$F$_6$ being
almost free to rotate, in the sense that it undergoes only minor
geometrical changes, presents a rather small rotational barrier of
$\sim$0.61 kcal/mol, while Si$_2$Br$_6$ has a rotational barrier of
$\sim$~2.6 kcal/mol, consistent with its larger geometric changes.
The above results indicate that torsional potential barriers, that hinder
the interconversion between two staggered conformations, arise from
structural rearrangements induced by an interplay between steric
repulsion and hyperconjugation effects~\cite{pop}.

It is important to remark that at the B3LYP/6-31G* level Si$_2$Br$_6$
is predicted to be stable in the eclipsed configuration, with a rather
significant energy difference of 1.17~kcal/mol relative to the
staggered one.  MP2/6-31G{*} also yields a smaller energy for the
eclipsed configuration, but with a much smaller difference of only
0.150~kcal/mol. However, the MP2/6-311++G{*}{*} calculations agree
with the B3LYP/6-311++G{*}{*} results. Thus, it seems that
6-311++G{*}{*} is the minimum basis set required to correctly describe
the rotational behavior of Si$_2$Br$_6$.

\begin{table}

{\centering \begin{tabular}{|c|c|} \hline Molecule&
Rotational Barrier(kcal/mol)\\
\hline \hline
 Si$_2$H$_6$ &
0.9441\\
\hline Si$_2$F$_6$&
0.6096\\
\hline Si$_2$Cl$_6$&
1.6272\\
\hline Si$_2$Br$_6$&
2.5920\\
\hline
\end{tabular}\par}
\caption{Calculated rotational barrier at the B3LYP/6-311++G**
level for \protect Si$_2$H$_6$,\protect \protect
Si$_2$F$_6$,\protect \protect Si$_2$Cl$_6$\protect and \protect
Si$_2$Br$_6$\protect } \label{table:rot_barriers}
\end{table}

Table~\ref{table:rot_barriers} displays the B3LYP/6-311++G{*}{*}
rotational barriers we obtained.  Substitution of the hydrogens, by the
more electronegative fluor atoms, results in a lowering of the
potential barrier. The electronic population is now mainly localized
at the SiF$_3$ groups thus weakening the Si-Si torsional bond. The
values for Si$_2$H$_6$ and Si$_2$F$_6$ compare well with those
calculated by Cho {\it et al.}~\cite{cho97}. The experimental values
for the rotational barrier of Si$_2$H$_6$ are $\sim 1$ kcal/mol,
and for Si$_2$F$_6$ between 0.51-0.73~kcal/mol, according to early
electron diffraction measurements~\cite{cho97}.

On the other hand, substitution of the hydrogens by chlorine and
bromine atoms tends to keep the electronic population uniformly
distributed, and the observed increase of the potential barrier seems
to be related to steric hindrance between quite voluminous chemical
groups.

\section{Chemical Potential and Hardness}
\label{sec:chem_pot}

In DFT the chemical potential of a molecule is defined by the
derivative of the energy with respect to the number of electrons
$N$ at constant external potential $v(r)$:

\begin{equation}
\mu =\left( \frac {\partial E}{\partial N}\right)_{v(r)} \; ,
\end{equation}

\noindent where $E$ is the energy and $N$ the number of particles.
For a finite system this extrapolation takes the
form~\cite{parr88}

\begin{equation}
\mu \cong \frac {1}{2}  [E(N+1)-E(N-1)] \; .
\end{equation}

\noindent Moreover, following Koopmans' theorem~\cite{koopmans}, the
anion energy \( E(N+1) \) can be approximated by
$E(N+1)\approx E(N)+E_{\mbox{\footnotesize LUMO}}$, and the cation energy
$E(N-1)$, by $E(N-1)\approx E(N)-E_{\mbox{\footnotesize HOMO}}$, where
$E_{\mbox{\footnotesize LUMO}}$ and $E_{\mbox{\footnotesize HOMO}}$
are the energies of the Lowest Unoccupied and Highest Occupied
Molecular Orbital, respectively. Within this approximation

\begin{equation}
\mu \cong \frac 12 (E_{\mbox{\footnotesize
    LUMO}}+E_{\mbox{\footnotesize HOMO}}) \; .
\end{equation}

Another relevant characteristic property we want to probe is the
chemical hardness $\eta$, defined as

\begin{equation}
 \eta =\frac{1}{2}\left(\frac{\partial ^{2}E}{\partial  N^{2}}\right)_{v(r)} \; ,
\end{equation}

\noindent which can be approximated by a finite difference as
follows:

\begin{equation}
 \eta \cong \frac 12 [E(N+1)+E(N-1)-2E(N)] \; ,
\end{equation}

\noindent which in terms of the HOMO--LUMO energies reads

\begin{equation}
\eta \cong \frac 12 [E_{\mbox{\footnotesize LUMO}} -
E_{\mbox{\footnotesize HOMO}}] \; .
\end{equation}

Nevertheless it is important to notice that the actual changes in
the torsional energy must include the geometrical changes induced
by the removal or addition of electrons. This geometrical
relaxation may include symmetry changes, such as the rotations
around the Si-Si axis we study in this paper. Thus, the
significance of \( \eta \) and \( \mu \) as calculated above,
whether with the HOMO--LUMO approximation or with the unrelaxed
(or constraint relaxed) cation and anion energies, is not
completely accurate.

\begin{figure}

  {\centering
    \resizebox*{!}{7cm}{\rotatebox{270}{\includegraphics{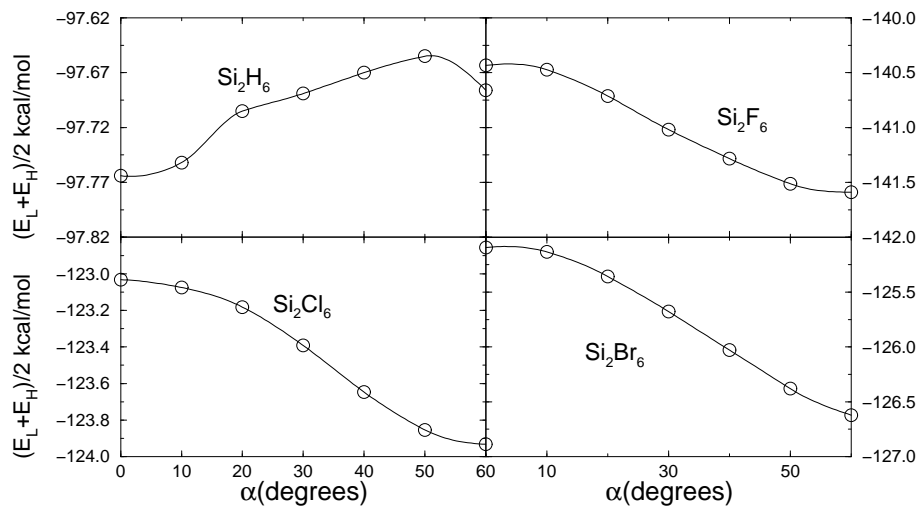}}}
    \par}
\caption{Chemical potential in the HOMO--LUMO approximation}
\label{fig:chem_pot}
\end{figure}

\begin{figure}

{\centering
\resizebox*{!}{7cm}{\rotatebox{270}{\includegraphics{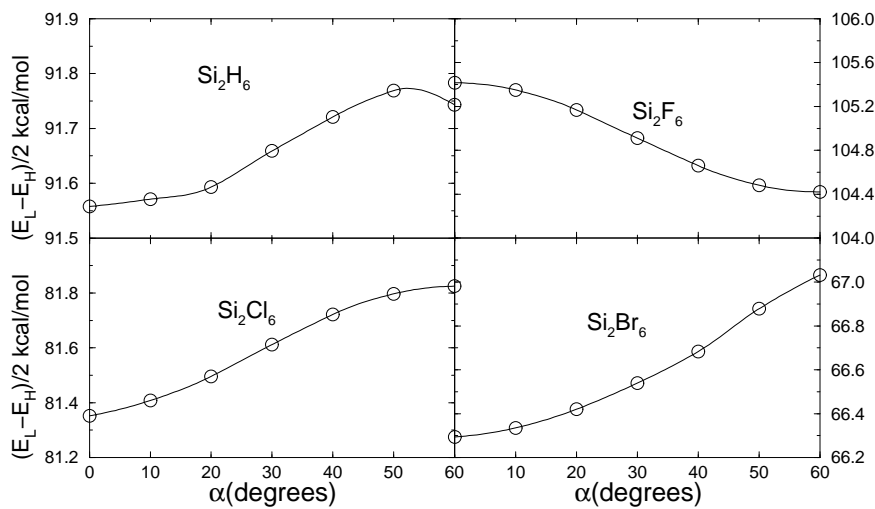}}}
\par} \caption{Chemical hardness in the HOMO--LUMO approximation.}
\label{fig:chem_hardness}
\end{figure}

Figs.~\ref{fig:chem_pot} and \ref{fig:chem_hardness} display the
chemical potential $\mu$ and chemical hardness $\eta$, respectively,
as a function of the torsion angle $\alpha$, in the HOMO--LUMO
approximation. An appreciably difference of the $\mu$ values between
two reference conformations implies that an electronic rearrangement,
with some charge transfer from the higher towards the lower $\mu$
conformation, will take place. By inspection of
Fig.~\ref{fig:chem_pot} we observe that the chemical potential versus
$\alpha$ profiles for all molecules, except Si$_2$H$_6$, display a
variation of $\mu \sim1$~kcal/mol as $\alpha $ varies by 60$^{\circ}$,
always opposite in sign to the relative to the energy variation
displayed in Fig.~\ref{fig:chem_hardness}.  This is an indication that
torsion implies a rearrangement of the electronic density. In
contrast, for Si$_2$H$_6$, the chemical potential remains quite
constant over the range $0\ge \alpha \ge 60^{\circ}$, with $\Delta \mu
\sim 0.10$~kcal/mol.

Further inspection of Fig.~\ref{fig:chem_hardness} reveals that the
overall hardness changes are quite small, ranging from $\sim
0.2$~kcal/mol for Si$_2$H$_6$ to $\sim1$~kcal/mol for Si$_2$F$_6$.
Si$_2$H$_6$, Si$_2$Cl$_6$ and Si$_2$Br$_6$ are chemically hardest in
the eclipsed conformation, while Si$_2$F$_6$ is hardest in the
staggered conformation. It is interesting to mention that the same
trends for the chemical hardness are predicted both by the
cation-anion energies, at the B3LYP/6-311++G{*}{*} level, and by the
HOMO--LUMO approximation at MP2/6-311++G{*}{*} level, as seen in
Table~\ref{table:chem_hardness}, where the numerical values of the
hardness in the staggered and eclipsed conformations are listed.

\begin{table}
\begin{center}
\begin{tabular}{|c|c|c|c|}
\hline Molecule& HOMO--LUMO, B3LYP& ANION CATION, B3LYP&
HOMO--LUMO, MP2\\
\hline \hline
 Si$_2$H$_6$ &
0.185& 0.308&
0.52\\
\hline Si$_2$F$_6$& -0.995& -0.856&
-0.90\\
\hline Si$_2$Cl$_6$& 0.473& 0.634&
\\
\hline Si$_2$Br$_6$& 0.738& 1.071&
1.39\\
\hline
\end{tabular}
\caption{Change in chemical hardness \protect\( \Delta \eta =
  \eta_{e}-\eta _{s}\protect \) in kcal/mol.}
\label{table:chem_hardness}
\end{center}
\end{table}

It is noticed that the PMH is verified only for Si$_2$F$_6$, while
Si$_2$H$_6$, Si$_2$Cl$_6$ and Si$_2$Br$_6$ present hardness profiles
obeying the same trend as their energy profiles. According to the PMH,
the hardness profile of Si$_2$F$_6$ displays a maximum at the stable
staggered conformation and a minimum at the unstable eclipsed
conformation. We want to emphasize the complementary behavior of
energy and hardness: whereas for Si$_2$F$_6$ the almost free internal
rotation does not allow to distinguish the energetically most
favorable $\alpha$ value, the hardness profile allows this
characterization. In contrast, the hindered rotation in Si$_2$X$_6$
(X=H, Cl, Br) yields energetically distinguishable conformations, but
they cannot be characterized by the hardness profiles.

\section{Reactivity of Silanes}
\label{sec:reactivity}

\begin{figure}
{\centering
\resizebox*{1\textwidth}{!}{\includegraphics{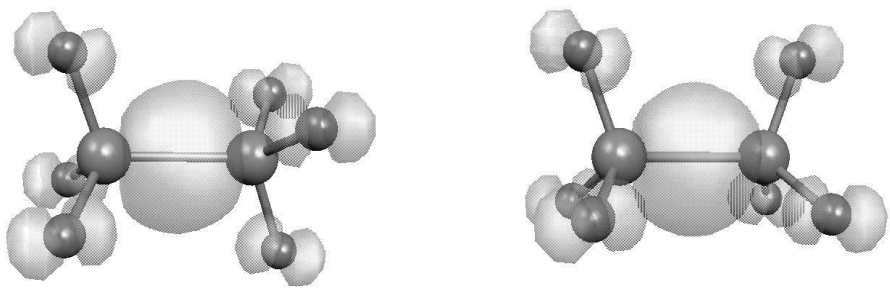}} \par}
\caption{HOMO for Si$_2$H$_6$. Left panel: staggered
configuration. Right panel: eclipsed configuration}
\label{fig:homo}
\end{figure}

The reactivity of these systems, induced by the internal rotation,
cannot be rationalized in terms of the profiles of $\mu$ and $\eta$
alone, due to their almost constant behavior as a function of
$\alpha$.  However, a different perspective of the electronic
structure and reactivity is provided by the LUMO and HOMO densities.
In Fig.~\ref{fig:homo} the HOMO of the Si$_2$H$_6$ molecule is shown
and we observe very similar orbitals to the other Si$_2$X$_6$
molecules we have considered. The majority of the orbital charge
accumulates on the Si-Si bond with some contribution on the hydrogens,
and with a bond of clear $\pi$-character. Moreover, there is little
difference between the HOMO staggered and eclipsed charge
distributions, indicating that the effect of the torsional motion on an
electrophilic attack is negligible.

\vspace{0.3cm}

\begin{figure}
\centering
\resizebox*{1\textwidth}{!}{\includegraphics{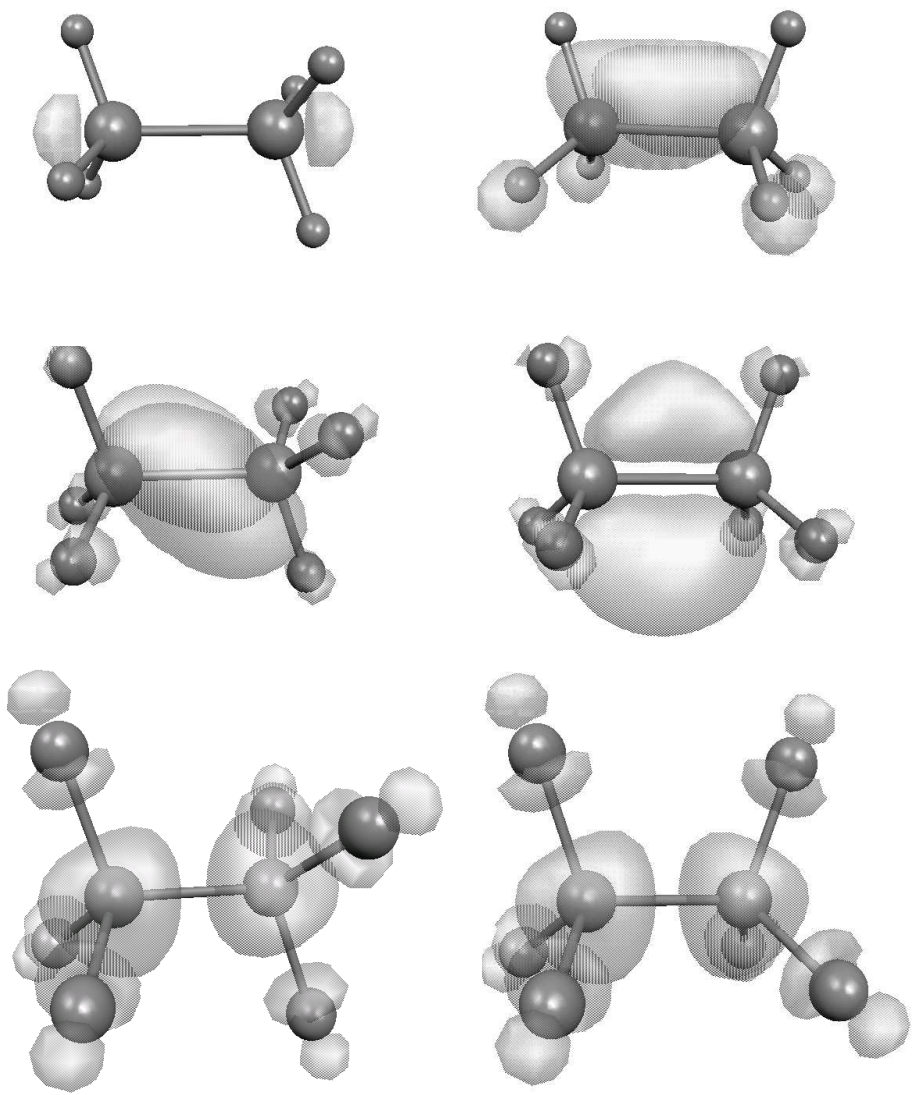}}
\vspace{0.3cm} \caption{LUMO for the different molecules
considered. Left panel: staggered configuration. Right panel:
eclipsed configuration. From top to bottom, Si$_2$H$_6$,
Si$_2$F$_6$ and Si$_2$Cl$_6$.} \label{fig:lumo}
\end{figure}

The LUMO electronic structure, as illustrated in Fig.~\ref{fig:lumo},
is completely different. For Si$_2$H$_6$, in the lowest energy
(staggered) configuration, the charge density is delocalized on the
sylil groups.  Moreover, inspection of Figs.~\ref{fig:homo} and
\ref{fig:lumo}, shows that the largest overlap between HOMO and LUMO
orbitals occurs for Si$_2$F$_6$, which suggests that in the staggered
configuration this molecule has the strongest hyperconjugative
effects~\cite{pop}.  On the contrary, in the eclipsed conformation the
charge is delocalized on the Si-Si bond, with a $\pi$-antibond
character. This indicates that a nucleophilic attack on Si$_2$H$_6$
may present different specific mechanisms, as a consequence of the low
torsion barrier. In the Si$_2$F$_6$ staggered configuration the
delocalization process is different: the charge is delocalized on the
Si-Si bond, with antibond character, and is symmetric around the Si-Si
bond, but with an asymmetry in the direction of the sylil groups.
Instead, in the eclipsed conformation, the charge is again delocalized
on the Si-Si bond with antibond character, but with some preferential
charge on the sylil side when viewed in a plane with four hydrogen
atoms. In addition, the delocalization volume is larger in the
staggered conformation. 

Finally, we consider the Si$_2$Cl$_6$ molecule (Si$_2$Br$_6$ behaves
similarly), which in its staggered configuration has the charge
localized on the Si atoms, with a very clear $\sigma$-character, but
with some asymmetry in the sylil group directions. This is similar to
the eclipsed configuration, where the charge distributes in much the
same way, except in that it is completely symmetric around the Si-Si
bond.  The above results suggest that the low, but significant,
barriers that hinder internal rotation may induce different specific
nucleophilic attack mechanisms.

\vspace{0.3cm}

\section{Concluding Remarks}
\label{sec:concl}

We have performed DFT calculations on Si$_2$H$_6$, Si$_2$F$_6$,
Si$_2$Cl$_6$, and Si$_2$Br$_6$ of the evolution of the electronic
energies, chemical hardness and chemical potentials as a function of
torsion angle. For all these molecules at the DFT-B3LYP/6-311++G**
level, the staggered conformation is predicted to be the most stable
one. Moreover, except for Si$_2$F$_6$, it is softer than the
eclipsed configuration due to a different charge delocalization at the
LUMO orbital.

Low, but significant, energy barriers hinder internal rotation.  For
Si$_2$H$_6$ the chemical potential and hardness remains quite constant
during the torsion process, while the other molecules show different
degrees of electronic density rearrangement as a function of the
torsion angle. However, it was not possible to characterize precisely
the reactivity behavior just on the basis of the chemical potential
and hardness profiles.

The qualitative analysis of the frontier orbitals shows that for the
Si$_2$X$_6$ series there is little difference between the HOMO
staggered and eclipsed charge distributions. This indicates that the
effect of the torsional motion on an electrophilic attack is
negligible. In contrast, the low but significant barriers that hinder
internal rotation may induce different nucleophilic attack mechanisms.

\end{document}